# Strategic Investment Decision Making for Value Creation in Energy Transition: A Reinforcement Learning Approach

Yasaman Cheraghi[1]*, Reidar B. Bratvold[1], Aojie Hong[2], Ressi B. Muhammad[1], and Sergey Alyaev[3]

[1]Department of Energy and Petroleum Engineering, University of Stavanger, Norway

[2]Independent Researcher, Stavanger, Norway

[3]NORCE Norwegian Research Centre, Bergen, Norway

# Abstract

The global challenge of climate change has driven significant steps to reduce $CO_2$ emissions, guided by international agreements like the Paris Agreement of 2015, which set targets for the transition from fossil fuels to renewable sources. Acting too slowly could result in future losses and reputational damage, while moving too quickly could jeopardize shareholder value due to the marginal profitability or potential losses due to technology immaturity of many renewable projects. To navigate this complex transition, energy companies must adopt Sequential Decision Making (SDM) strategies to maximize value creation from decision flexibility under uncertainties. However, achieving this goal is not trivial. This necessitates the use of decision-analysis and optimization methods that deliver robust decision insights without prohibitive computational costs or delays.

To support this, we developed a custom simulation environment to model the dynamic energy landscape up to 2050. Building on this, we designed a multi-criteria SDM framework that explores various decision strategies related to different portfolios for allocating funds across three sectors: oil & gas, renewables, and $CO_2$ reduction. In the context of long-term investments, such a framework also helps energy companies better understand their values and the trade-offs between them. It aims to maximize value during the transition while accounting for uncertainties in productions, energy prices, and costs. This framework has three objectives: maximizing profit, minimizing $CO_2$ social costs, and enhancing competitive advantage in the renewable energy sector.

This research evaluates the use of Reinforcement Learning (RL) to identify optimal investment policies within the defined SDM framework. The agent's sequential decisions shape a virtual dynamic environment by influencing key variables such as oil and gas production, renewable energy output, $CO_2$ emissions, and revenues. In turn, the environment responds with updated state variables and immediate feedback based on the three defined objectives. Through repeated interaction, the RL algorithm explores the state space and learns an optimal policy under uncertainty. We benchmark the RL strategy against a set of manually defined baseline policies and find it consistently outperforms them in adaptability and long-term value creation.

# Introduction

Human activities are now the primary driver of Earth's atmospheric composition changes, leading to significant changes in the climate system (Hansen et al., 2013) . The global trend of urbanization with escalating energy consumption and associated greenhouse gas (GHG) emissions, has contributed to a significant rise in atmospheric GHG levels (Kumar, 2018). The urgent need to mitigate the environmental impacts of climate change has led nations to establish global policies and agreements targeting GHG emissions, such as the 2015 Paris Agreement (Erickson & Brase, 2019).

The energy sector is the main contributor to GHG emissions, primarily through the combustion of fossil fuels for electricity and heat generation (IEA, 2023). This sector is undergoing a transformative shift to reduce reliance on fossil fuels and transition towards renewable and sustainable energy resources such as wind, solar, hydro, and geothermal power (Alagoz & Alghawi, 2023; Gielen et al., 2019).

However, such a transition is not trivial and causes severe challenges for energy companies, particularly in determining the optimal timing and scale of investments in renewable energy resources, as well as in balancing conflicting values within their transition strategies. These challenges arise primarily from the inherent uncertainty surrounding future technological advancements, energy demand projections, and market conditions for renewable energy resources (IPCC, 2011; Jacobsson & Bergek, 2011; Kern & Markard, 2016).

If energy companies delay transitioning to renewable energy sources, they risk losing future profit, competitiveness, and reputation in such a transition. On a larger scale, as noted by the International Energy Agency (IEA), delaying efforts to reduce $CO_2$ emissions may lock in unsustainable emission levels (IEA, 2009) and Carbon Capture and Storage (CCS) is widely recognized as a critical component for achieving net-zero emissions (IEA, 2023). However, moving quickly into investments in renewable energy may jeopardize shareholder value due to uncertain profitability and technology challenges in many renewable projects, such as weather dependence issues with solar and wind energy, regulatory uncertainties, and the need for substantial upfront investment without immediate returns (Miller et al., 2013; Painuly, 2001; Sovacool, 2016). Giddens has introduced the term "climate paradox" to describe the dilemma where humanity may recognize the urgent need to transition to low-carbon energy sources only after reaching a point where irreversible consequences of climate change are already unfolding (Giddens, 2009). Energy companies face a similar dilemma as they must decide on the optimal time and scale of this transition. This dilemma was the motivation of the current study.

This research has three main objectives. First, it conceptualizes the energy transition as a Sequential Decision-Making (SDM) process under uncertainty to evaluate different planning strategies and highlight the benefits of this dynamic approach compared to the static models commonly used in long-term energy planning. Second, it develops a custom, open-access simulation environment that models the evolving energy landscape and can be used as a valuable tool for future research on long-term investment strategies in the energy transition. Third, it applies Reinforcement Learning (RL) within the custom simulation environment as a state-of-the-art approach for solving complex SDM problems, with the goal of identifying optimal timing and scaling strategies for the transition. By incorporating RL techniques, the study aims to enhance decision-making capabilities in the energy sector long term planning and advance methodological tools for addressing its evolving challenges.

Portfolio management involves the continuous or periodic process of reallocating funds across various financial investment opportunities, to maximize returns while minimizing risks (Markowitz, 2008). This process is particularly critical during the energy transition, where strategic fund allocation and prompt decision-making are essential for balancing investments across various sectors. As the global energy landscape shifts towards more sustainable practices, portfolio management helps determine how and when to reduce dependency on oil and gas while scaling up investments in cleaner, renewable alternatives like wind, solar, and hydroelectric power. Despite such high importance, there remains a limited body of research specifically focusing on long-term portfolio management within the context of the energy transition.

The studies conducted so far in the context of portfolio management within energy sector often focus on specific segments, such as renewable energy integration, grid management, or risk assessment related to specific technologies (Bender et al., 2020; Bistline, 2016; Bressan et al., 2022; Ender et al., 2010; Semmler et al., 2024). For example, in a study by Hocine et al., the selection of a renewable energy portfolio under uncertainty was addressed using the multi-segment fuzzy goal programming (MS-FGP) method. The authors extended traditional fuzzy goal programming to handle the diverse and often conflicting criteria associated with renewable energy projects. They applied their approach to a real-world scenario, optimizing a portfolio of solar, wind, biomass, and tidal energy sources (Hocine et al., 2018). In another study conducted by Wu et al., a new energy portfolio management system tailored to the complex landscape of China's energy projects is developed. This system leverages project portfolio management technology to address the shortcomings of traditional project management in handling regulated energy projects (Wu et al., 2012). Semmler (2020) examines how short-termism affects investment decisions, particularly in portfolios that include green assets. The study models a dynamic portfolio choice problem involving a volatile asset and a risk-free asset, analyzing how different discount rates and investment horizons influence outcomes in both deterministic and stochastic settings.

The way we approach portfolio management in the context of the energy transition is novel and takes a broader, more integrated perspective. We propose a multi-criteria SDM framework designed to navigate the evolving energy landscape through 2050. This framework is built on a dynamic yet simplified simulation environment, which will be described in detail in the next section. Within this environment, an energy company must make sequential decisions about how to allocate funds across projects and when to invest, to optimize multiple objectives, such as maximizing profit, minimizing social costs of $CO_2$, and enhancing competitive advantage in renewables.

We use RL as a powerful tool to find the optimal investment strategy within our SDM framework. The energy company acts as the decision-maker, making investment decisions that influence the behavior of the environment. The environment, in turn, responds to these decisions, by updating key variables and uncertainties and by providing feedback on the immediate outcome. Through repeated interactions with the simulation environment, the RL algorithm navigates the broad space of uncertain and deterministic variables. By systematically exploring and exploiting alternative decisions, it learns an optimal investment strategy to maximize long-term value.

The use of RL for portfolio management is not a novel concept, with several studies applying RL algorithms across diverse portfolios (Betancourt & Chen, 2021; Filos, 2019; Hu & Lin, 2019; Jiang & Liang, 2017; Jiang et al., 2017; Liang et al., 2018; Wu et al., 2021). However, our literature review revealed a gap in its application within the specific context of strategic decision-making for energy transition investment strategies.

Our results demonstrate that the RL-based investment strategy dynamically reallocates funds over time, gradually reducing investments in oil and gas while increasing allocations to renewables and $CO_2$ reduction. Under this strategy, oil production peaks at about 109 million barrels of oil equivalent around 2033 and then declines to roughly 26 million by 2050, while renewable electricity generation expands to about 27,000 GWh over the same period. As one of the constraints defined in the model, the RL policy achieves near-zero $CO_2$ emissions by 2050, with a consistent downward trend starting around 2033. However, this ambitious reduction comes at a cost: annual cash flow declines through the 2030s to a minimum in the early 2040s before partially recovering as renewable capacity matures. This highlights the critical challenge of the energy transition, that achieving net-zero emissions requires accepting substantial financial sacrifices during the transition.

Compared to several benchmark strategies that will be described in detail in the Results section, the RL policy achieves higher long-term value.

The remainder of this paper is organized as follows. We begin by presenting the simulation framework and defining its variables. Next, we introduce the theoretical background, formulating the problem as a Markov Decision Process and explaining the principles of RL, especially the Deep Q-Network (DQN) algorithm. We then describe how the investment decision problem is structured within the RL framework, detailing the state and action spaces, reward function, and the neural network architecture. This is followed by the implementation details, training setup, and results of the optimal investment strategy, along with comparisons to benchmark policies. Finally, we conclude the study with a discussion of key findings, limitations, and suggestions for future research.

# Simulation Environment

In this study, we develop a decision-analytic framework to model a conceptual energy company in the energy transition over a 25-year period within a dynamically simulated environment. It is important to note that this framework is not based on an existing energy-transition model or on data from any specific company; rather, we define the overall structure, components, and their interactions ourselves, while grounding all assumptions in the literature, public reports, and logical trends representative of a typical energy company and the real-world conditions. A stochastic environment is designed to incorporate key components such as oil and gas production, $CO_2$ emissions and associated costs, renewable energy production, energy prices, and operational expenses, as illustrated in the decision diagram below (Figure 1).

## Decisions

The decision variables in this framework are the allocation of investment across the three sectors each year. These allocations are expressed as fractions of the total available investment budget in that year and sum to one. The available investment budget evolves dynamically over time based on the financial performance of the company. The details of this budgeting mechanism are described in the following sections.

The time horizon of this framework is set to 25 years, modeling the energy transition from 2025 to 2050 which aligns with the Paris Agreement's target for net-zero emissions. Each year, investment decisions are made by the company that shape the evolving energy landscape within the environment. Decision nodes are shown by rectangles in the diagram.

The first decision is taken in 2025 based on the initial conditions of the environment, and each decision influences the state of the following year. The model therefore comprises 25 sequential decisions, taken from 2025 to 2049, with the resulting outcomes realized between 2026 and 2050. Accordingly, all result figures report outcomes over 2026–2050, while the initial conditions listed in this section correspond to the state in 2025.

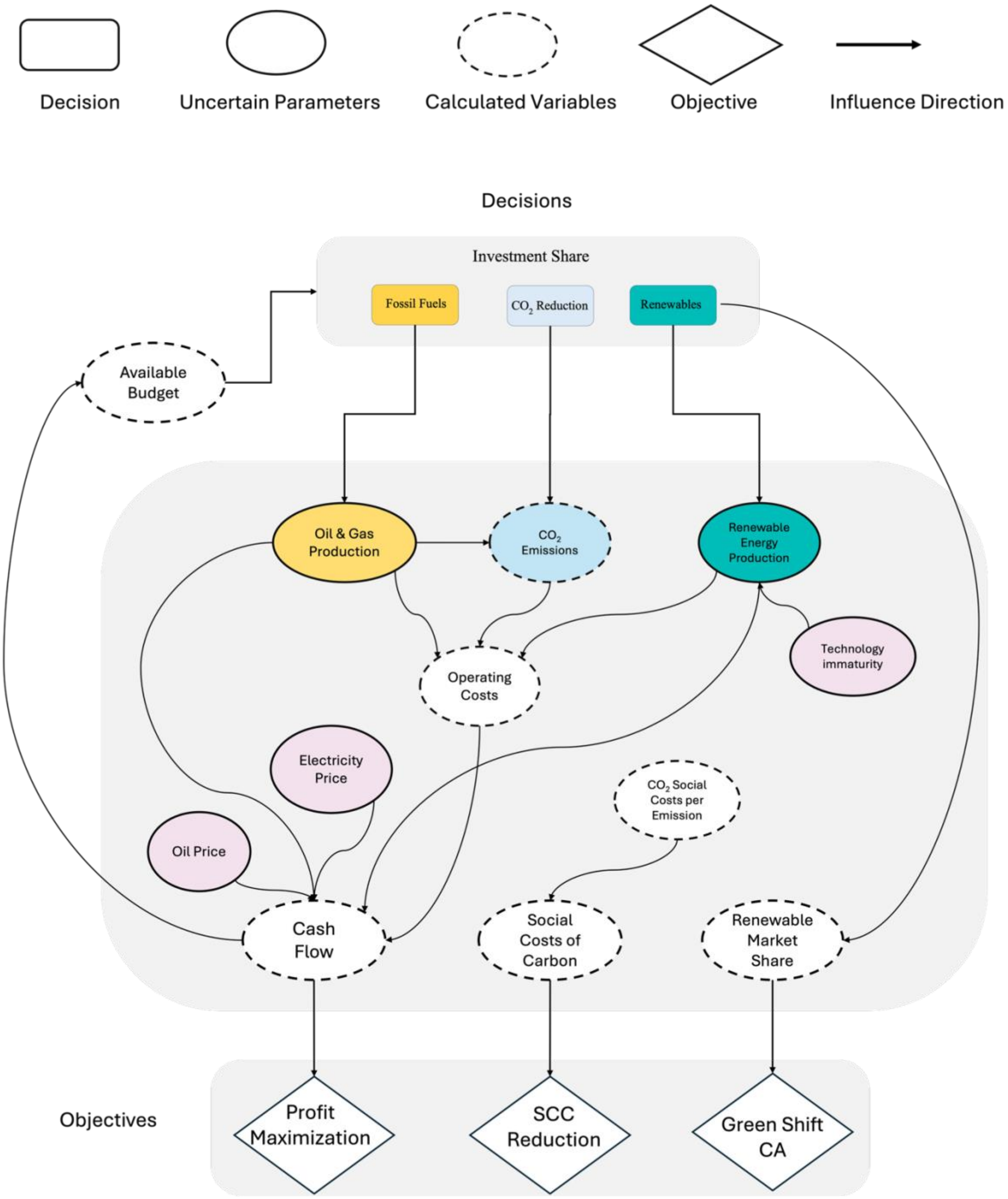

**Figure 1.** Influence diagram of the simulation environment.

## Environment

The environment developed in this study consists of key variables essential to the energy transition and the interrelations between them. While this environment is a simplified representation of the real world, it provides a general and structured decision-supporting framework, with representative assumptions as an example, to explore and evaluate different investment strategies under uncertainty. Our developed environment is publicly available at: https://github.com/YasCheraghi/Energy-Transition-RL_Third_Paper_PhD for research, aiming to support reproducibility, facilitate benchmarking, and encourage extension of the environment and further studies on energy transition investment strategies.

The variables in this environment can be categorized into two main groups. The first group includes variables whose evolution is influenced, directly or indirectly, by investment decisions. This group comprises stochastic and deterministic variables such as oil and gas production, $CO_2$ emission volumes, electricity production from renewable resources, operational costs, $CO_2$ social costs, cash flow, and renewable market share. In the influence diagram (Figure 1), primary variables affected by decisions are shown as yellow, blue, and green solid ovals, with colors corresponding to the investment share that influences them. The derived variables calculated based on defined models and other variables, are shown as dashed ovals. The second category comprises exogenous variables that evolve stochastically over time and are not affected by investment decisions or other model components. These include oil prices, electricity prices, and the risk of technological immaturity in renewable energy. These variables are represented as solid ovals in a distinct color, pink, to indicate their independence from decision nodes.

## Group 1: Decision-dependent Variables

### Oil and Gas Production

For oil and gas production, we model annual production based on investment level, reserve change, and production growth. Initial production is 100 million barrels of oil equivalent (BOE) per year, with initial total recoverable reserves ($Res_{total,\ init}$) of 10,000 million BOE and an initial annual decline rate ($D_{init}$) of 9%. The initial reserves can be expanded at a rate of $ExpRate$ = 0.05 BOE per USD invested. We assume a fraction of $Share_{ResExp}$ = 0.3 of the investments in oil & gas sector ($Inv_{oil,i}$) is allocated to reserve expansion, and the remaining is directed toward production capacity growth.

The decline rate is dynamically adjusted in proportion to the ratio of cumulative production to total reserves ($Prod_{CumOil,i}$ $/Res_{total,i}$). The updated production level in year $i$, $Prod_{oil,i}$, is calculated based on the previous year's production ($Prod_{oil,i-1}$), declining at the updated decline rate ($D_i$) and the production increase resulting from new investments. The updated production is also constrained by the remaining reserves to ensure physical consistency. The initial yearly production of a new well ($Prod_{ave}$ ) is modeled as a stochastic variable sampled from a normal distribution with a mean of 1.46 million BOE and a standard deviation of 0.146 million BOE (10% of the mean). The sampled initial yearly production per new well is then multiplied by the number of drilled new wells ($N_{new}$), where the well cost is considered as $Cost_{well}$ = 100 million USD. The following equations describe the oil production update process:

$$Prod_{oil,i} = Prod_{oil,i-1} * (1 - D_i) + Prod_{ave}^{N_{new,i}} \quad \text{Eq. 1}$$

$$Prod_{ave}^{N_{new,i}} \sim \mathcal{N}(mean = N_{new,i} * 1.46, variance = N_{new,i} * 0.146^2)$$

$$N_{new,i} = Inv_{oil,i} * (1 - Share_{ResExp}) / Cost_{well}$$

$$D_i = D_{init} * (1 + Prod_{CumOil,i-1} / Res_{total,i}) \quad \text{Eq. 2}$$

$$Res_{total,i} = Res_{total,i-1} + Inv_{oil,i} * Share_{ResExp} * ExpRate \quad \text{Eq. 3}$$

## Renewable Energy Production

For renewable energy production, we model annual installed capacity and electricity production based on investment levels, efficiency improvements, and infrastructure degradation. The model considers a portfolio of wind power, solar, and hydropower, recognizing that the share of each renewable resource varies significantly across countries due to regional climatic differences. In this study, wind power is assumed to dominate the renewable mix. The initial installed capacity is assumed to be 2 GW, with a starting electricity production of 2000 GWh per year.

Investment in renewables leads to additional capacity installation and increased electricity generation. However, the model also incorporates efficiency improvements over time, risks associated with technological immaturity ($TechImmaturityRisk_i$) which will be explained in a later section, and the gradual degradation of renewable infrastructure. The annual updates to installed capacity and electricity production are computed as:

$$Capacity_{rnw,i} = Capacity_{rnw,i-1} + Inv_{rnw,i} * CapPerInvest \quad \text{Eq. 4}$$

$$\eta_i = \eta_{init} + i * EffGrowth \quad \text{Eq. 5}$$

$$Prod_{rnw,i} = Capacity_{rnw,i} * (1 - DegRate) * \eta_i * h * TechImmaturityRisk_i \quad \text{Eq. 6}$$

In these formulas, $Capacity_{rnw,i}$ is the updated capacity in year *i,* calculated by a capacity factor assumed to be $CapPerInvest$ = 1/2000 GW per million USD of the investment allocated to renewables ($Inv_{rnw,i}$ ). Electricity production ($Prod_{rnw,i}$ ) is updated based on the additional installed capacity.

Electricity production is scaled by an efficiency factor ( $\eta_i$), which reflects annual improvements in renewable energy generation technology. The initial efficiency is assumed to be $\eta_{init}$ = 25%, based on a weighted average of the renewable energy resource portfolio. To account for technological progress over time, this efficiency is assumed to increase linearly at a rate of $EffGrowth$ = 1% per year. The efficiency factor is applied to the updated capacity in each year. Infrastructure degradation is represented by an aggregate degradation rate applied to annual production at an assumed rate of $DegRate$= 1.35 %.

The number of full-load operational hours, $h$, varies significantly due to climatic conditions, weather fluctuations, and other environmental factors. We model it as an uncertain variable, sampled from a normal distribution with a mean of 3,550 full-load hours per year, based on a weighted average of the renewable energy resource portfolio, and a standard deviation of 10% of the mean.

### $CO_2$ Emission Volumes

The $CO_2$ storage model quantifies the amount of $CO_2$ capture and storage (CCS) through investments in Carbon Reduction methods. The volume stored in year $i$, ($CCS_i$) is proportional to the investment allocated to this sector in that year ( $Inv_{CO_2,i}$ ) via a CCS efficiency factor assumed to be $StorageEfficiency$ = 0.0125 tons of $CO_2$ per invested USD:

$$CCS_i = Inv_{CO_2,i} * StorageEfficiency \qquad \text{Eq. 7}$$

For $CO_2$ emission modeling, we regress on publicly available data from Equinor's Climate Tables (Equinor, 2024). The dataset includes annual oil equivalent productions and $CO_2$ emissions from 2016 to 2023. Based on this model, approximately $EmissionsPerBOE$= 0.3 million tons of $CO_2$ are emitted per million BOE produced. In our model, we consider only $CO_2$ emissions associated with fossil fuels consumption, and minor emissions from renewable energy development projects are ignored because their contribution to total emissions is negligible compared to fossil fuels.

The net annual $CO_2$ emissions are the annual amount of $CO_2$ generated from hydrocarbon production minus the annual amount of $CO_2$ captured and stored ($CCS_i$). Net emissions cannot be negative and are floored at zero if capture and storage exceed emissions.

$$Emissions_{CO_2,i} = Max\,(EmissionsPerBOE * Prod_{oil,i} - CCS_i\,, 0) \qquad \text{Eq. 8}$$

### Operating Costs

The operating costs for oil & gas production ($OpCost_{oil,i}$) , renewable electricity production ($OpCost_{rnw,i}$) and CCS projects ($OpCost_{CO_2,i}$) start at 15 USD per BOE, 0.015 USD per kWh, and 50 USD per ton CCS, respectively. These costs change annually by a fixed rate of $AnnualCostDrift$ = 5% for oil & gas production, 1% for CCS projects and -1% for renewable energy production projects.

$$OpCost_{project,i} = OpCost_{project,0} * (1 + AnnualCostDrift)^i \qquad \text{Eq. 9}$$

### Available Budget

The available investment budget is modeled as a dynamic variable that evolves with the company's financial performance. A base investment level ($B_{base}$ = 1 Billion USD) ensures continuous activity, while a fraction (φ = 25%) of the smoothed operating cash flow is reinvested to expand future investment capacity. The smoothed operating cash flow, ${CF^*}_i$, incorporates memory of past performance through exponential smoothing with persistence $\rho$ = 0.7 (Eq. 10a); only positive operating cash flow ($CF_i^{OP}$) accumulates, so that loss years do not erode reinvestment capacity. To avoid unrealistic fluctuations, the budget does not jump to its target level but adjusts gradually toward it, closing a fraction (κ = 0.25) of the gap each year (Eq. 10b), and is bounded between $B_{min}$= 500 and $B_{max}$ = 10,000 million USD.

$$ {CF^*}_i = \rho \, . {CF^*}_{i-1} + (1-\rho). \max\left( 0, \; CF_i^{OP} \right) \qquad \text{Eq. 10a} $$

$$ B_i = clip\left( (1-\kappa). B_{i-1} + \kappa \, . (B_{base} + \varphi . {CF^*}_i), \; B_{\min}, B_{\max} \right) \qquad \text{Eq. 10b} $$

## Group 2: Independent Variables

### Oil & Electricity Price Model

Oil and electricity prices are modeled, as independent of the decision variables, following Geometric Brownian Motion (GBM) processes. GBM is a standard representation of long-term commodity price uncertainty in investment valuation (Semmler, 2020; Smith & McCardle, 1999) and requires three parameters: the initial value, the volatility, and the drift rate. The assumed volatilities are deliberately modest relative to historical spot-price volatility: they represent uncertainty in smoothed annual average prices at the model's yearly decision frequency rather than short-term market variability. In this model, the initial oil price is assumed to be 85 USD per barrel, while the initial electricity price is 0.1 million USD per GWh. Both prices are assumed to have the same volatility of 3% but different drift rates, -0.3% for oil price and 3% for electricity price. The price in each year ($Price_i$ ) is calculated using the GBM model as:

$$ Price_i = Price_{i-1} * e^{(\mu - \frac{\sigma^2}{2}) + \sigma * z} \qquad \text{Eq. 11} $$

where $\mu$ is the drift rate, $\sigma$ is the volatility, and $z$ is a standard normal random variable.

### Technology Immaturity Risk Model

We model this risk as an annual Bernoulli underperformance event: in year i an event occurs with probability $p\ (i)$, and when it occurs the year's renewable electricity output is reduced by a severity factor $s$ = 80% (Eq. 6). The probability is high in early years and falls to zero late in the horizon (Eq. 12), reflecting external technology maturation outside the company's control.

$$p\ (i) = \begin{cases} 0.7 & i \le \frac{T}{5} \\ 0.2 & \frac{T}{5} < \ i < \ 4\frac{T}{5} \\ 0 & 4\frac{T}{5} \le i \end{cases} \qquad \text{Eq. 12}$$

where $T$ is the total number of years which is 25 in this work and $i$ is the index of year and the $TechImmaturityRisk_{\,i}$ used in Eq. 6 will be defined as $(1 - \ X\ (i).s)$, where $X\ (i) = \ Bernoulli(p(i))$ indicating whether an underperformance event occurs in year $i$.

# Objectives and Constraints

This SDM framework defines three key objectives that are essential for energy companies and their shareholders in navigating the energy transition: maximizing profit, reducing social costs of carbon, and enhancing Competitive Advantage (CA) in green shift. These objectives span monetary, environmental, and strategic dimensions, making direct comparison challenging. To address this, we convert all three objective values into a common monetary value, enabling their integration into a single maximization objective.

## Maximizing Profit

Maximizing profit is a fundamental objective for any energy company, as it directly impacts financial sustainability, shareholder value, and new investment capacity. In the context of the energy transition, maintaining profitability is particularly challenging due to market uncertainties, fluctuating energy prices, and evolving regulatory frameworks. Ensuring financial sustainability enables companies to reinvest in renewable energy projects, technological advancements, and carbon reduction strategies. To quantitatively assess profitability, we use the Net Present Value (NPV) as the measure of this objective. NPV accounts for revenues, operational costs, and discounting of future cash flows. The net cash flow (NCF) in each year $i$ and the NPV of the NCF are calculated as:

$$NCF_i = \ Prod_{oil,i} * Price_{oil,i} \ + \ Prod_{rnw,i} * Price_{electricity,i} \ - \ (OpCost_{oil,i} * \ Prod_{oil,i} + \ \ OpCost_{rnw,i.} * Prod_{rnw,i} + \ \ OpCost_{co2,i} * \ \ CCS_{\,i}\,) \ - \ B_{\,i} \qquad \text{Eq. 13}$$

$$NPV_i = \frac{NCF_i}{(1 + r\ \ )^{i+1}} \qquad \text{Eq. 14}$$

where we assume the risk-free discount rate $r=5\%$.

## Reducing Social Costs of $CO_2$ (SCC)

Reducing GHG emissions is the core objective of the energy transition. In this study we particularly focus on $CO_2$ because $CO_2$ is the major composition of GHG. The social costs of carbon (SCC) quantify the economic damages per ton of $CO_2$ emissions, incorporating climate-related damages, productivity losses, and societal impacts. Accounting for SCC is essential for aligning investment strategies with long-term sustainability goals (Nordhaus, 1993).

We draw on insights from the Dynamic Integrated Climate-Economy (DICE) model developed by Nordhaus (2018; 1992) to estimate SCC. The model links carbon emissions, economic growth, and climate-induced damages, providing a structured approach to internalizing $CO_2$ costs. In line with values commonly reported in the integrated assessment literature (Rennert et al., 2022), we assume an initial value of \$70 per ton of $CO_2$ emission ($SCC_{0,per\ ton}$), increasing at a rate of $SCCgrowth$ = 2 % annually to reflect escalating climate-related damages, with a discount rate of $dr$ = 3% to account for the time value of money. Then the SCC in each year is calculated as:

$$SCC_i = Emissions_{CO_2,i} * SCC_{0,per\ ton} * (1 + SCCgrowth)^{i+1} * \frac{1}{(1 + dr)^{i+1}} \quad \text{Eq. 15}$$

We intentionally use a lower rate than the financial discount rate $r$ = 5%, following the convention that social discount rates, reflecting intergenerational equity, are below private rates (Rennert et al., 2022).

## Enhancing Competitive Advantage (CA)

Defining competitive advantage is particularly challenging, as it is a strategic objective rather than a directly measurable metric. In this framework, competitive advantage is represented by the company's market position within the global renewable energy industry, assumed to be worth \$1 trillion in the beginning. The company's market share in renewables serves as a proxy for its leadership and influence in the sector, with the timing of investment being a key determinant.

To maximize competitive advantage, the company must position itself as an Early Adopter, as described in Geoffrey Moore's "Crossing the Chasm" (1999). Early Adopters recognize the long-term potential of emerging technologies and invest proactively, securing first-mover advantages, technological expertise, and stronger market positioning before competitors enter the space. However, investing too soon carries risks, as technological immaturity and uncertain returns may lead to short-term financial losses. Successfully bridging the gap to the mainstream market requires demonstrating scalability, reliability, and cost-effectiveness to ensure long-term market leadership in the renewable energy transition.

To model competitive advantage, we assume a maximum base market share of 0.1% of the global renewable energy market. This base share is secured if the company invests early and acts as an Early Adopter, with a five-year window to establish its position. Within this period, the company must allocate more than 15% of its available investment budget to renewables in order to secure the base share. If this condition is met for fewer than five years, the base share is adjusted proportionally. This formulation is shown in Eq. 16, in which $n$ is the number of years the threshold is achieved.

To incorporate delayed effects in renewable development, we do not assign any immediate return to this objective during the first five years, unlike the other two objectives. This reflects the fact that renewable investments typically require several years before generating observable benefits.

However, early investment alone is not sufficient, and a continuous increase in renewable investment is also necessary to enhance competitive advantage. After the initial five-year period, competitive advantage evolves recursively as a state variable, depending on both its previous value and year-to-year changes in renewable investment share. If the renewable investment share increases by at least 3 percentage points over the previous year, CA grows by 5%; if it is maintained or increases by less than 3 points, CA grows by 1%; if it decreases, CA declines by 10% multiplied by the relative decline fraction (previous − current)/previous, capped at 1. Based on this approach, the CA reward for each year is calculated as:

$$Base\ Share = \min\left(\frac{n}{5}, 1\right) * Maximum\ Base\ Share \qquad \text{Eq. 16}$$

$$Growth\ Rate = \begin{cases} 5\% & \Delta_i \geq 3\% \\ 1\% & 0 \leq \Delta_i < 3\% \\ 10\% * (\Delta_i \ / \ Inv_{rnw,i-1}) & \Delta_i < 0 \end{cases} \qquad \text{Eq. 17}$$

$$CA_i = \begin{cases} 0 & i < 5 \\ Base\ Share * (1 + Growth\ Rate) & i = 5 \\ CA_{i-1} * (1 + Growth\ Rate) & i > 5 \end{cases} \qquad \text{Eq. 18}$$

In this study, CA objective is intended to capture strategic value that the cash-flow model does not represent; revenues from the company's own renewable production already enter the profit objective, whereas CA reflects the value of an established position in the renewable market: expertise, market access, and the ability to expand as the market grows, including beyond the modeled horizon. We acknowledge that market position and future sales are not fully separable, and that our market-share proxy is a simplification. Because CA represents a strategic position rather than a cash flow, it enters the reward as an undiscounted annual index.

## Net-Zero Emission Regulations

Beyond the three core objectives, we introduce regulatory constraints to align with net-zero emission policies. $CO_2$ emission reduction milestones are set at 10% by 2035, 45% by 2040, 75% by 2045 and 95% by 2050, reflecting common climate targets (United Nations, 2021). While these vary across countries and companies, they provide a structured pathway for the transition. The model evaluates whether $CO_2$ emissions have been reduced by the specified amount. Failure to meet these targets will add penalties to the reward.

## Smooth Transition Constraint

While sharp shifts in investments may seem optimal once uncertainties unfold, real-world transitions cannot happen instantly. Scaling up investments requires time to develop human resources, infrastructure, and supply chains, while scaling down involves gradual asset decommissioning and workforce reallocation. To limit significant changes in investment shares from one year to another year, we impose a penalty on abrupt investment shifts. Specifically, if the change in the allocated share of investments for any sector reaches or exceeds 0.15 within a single year, a penalty is applied.

# Theoretical Background

In this section, we formulate the investment decisions as a SDM process and frame and solve it within the RL framework. We begin by introducing Markov Decision Processes (MDPs), a fundamental mathematical framework for SDM. We then provide a brief overview of RL as a solution method for MDPs. Finally, we present the DQN algorithm used in this study, along with the evaluation metrics and implementation details of our approach.

## Markov Decision Processes

MDPs are a fundamental framework to model stochastic sequential decision-making (Puterman, 1990). In an MDP, a decision-making agent interacts with the environment over a sequence of discrete time steps ($t$). At each time step, based on the current state, $s \in S$, the agent selects an action $a \in A$ which causes the environment to transition to a new state $s'$ and return an immediate reward $r \in R$, measuring the effectiveness of the selected action. A transition function $p\,(s', r \mid s, a\,)$ defines the probability of the transition between states. The objective is to maximize cumulative long-term rewards.

A key assumption in MDPs is the Markov property, which states that the probability of reaching the next state $s'$ and receiving reward $r$ depends only on the current state $s$ and action $a$, independent of past states and actions. This property is particularly essential in SDM as it allows for the formulation of recursive relationships between states. Bellman Equation defines this recursive relationship between the values of two successive states (Bellman, 1966). The state-value function $V_{\pi}(\mathrm{s})$ is the expected cumulative reward an agent can obtain from state $s$ while following a fixed policy, $\pi(\, a \mid s)$, the probability of taking action $a$, in the state $s$.

Bellman Equation states the value of being in state $s$, $V_{\pi}(\mathrm{s})$, equals the immediate reward $r$ from transitioning to the next state, $s'$, plus the discounted value of the next state $\gamma V_{\pi}(\mathrm{s}')$. The discount factor $\gamma$ ranges from 0 to 1 and determines the relative importance of future rewards with $\gamma$ = 0 emphasizing immediate rewards and $\gamma$ approaching 1 emphasizing future rewards. Obtaining the expected value over all possible actions in the current state, as well as all potential next states and rewards, leads to the formulation of Bellman Equation for state-value function:

$$V_{\pi}(s) = \sum_{a \in A} \pi(\, a \mid s) \sum_{s' \in S, r \in R} p\,(s', r \mid s, a\,)\,[\, r + \gamma V_{\pi}(s')] \qquad \text{Eq. 19}$$

Bellman Equation can be extended to define the action-value function Q (s, a), which evaluates state-action pairs:

$$Q_{\pi}(\mathrm{s}, \mathrm{a}) = \sum_{s' \in S, r \in R} p\,(s', r \mid s, a\,) \left[ r + \gamma \sum_{a' \in A} \pi(\, a' \mid s')\, Q_{\pi}(s', a') \right] \qquad \text{Eq. 20}$$

Under an optimal policy, the agent always selects the action that maximizes the expected future rewards. Q-function for the optimal policy ( $Q_*$ ) is expressed by Bellman Optimality Equation:

$$Q_*(s, a) = \sum_{s' \in S, r \in R} p\,(s', r \mid s, a\,)\,[\, r + \gamma\, max_{a'} Q_*(s', a')\,] \qquad \text{Eq. 21}$$

Dynamic Programming (DP) is a collection of mathematical algorithms that are used to find the optimal policies for MDPs when the environment's dynamics are fully known (Sutton & Barto, 1998). DP operates by turning Bellman equations into update rules for computing value functions iteratively. DP faces two major limitations. First, its assumption of a complete knowledge of the environment, which is often impractical in real-world applications. The second limitation is the substantial computational expense known as the curse of dimensionality. As the number of state variables increases, the state space grows exponentially, making computations infeasible for large-scale problems.

## Reinforcement Learning

RL is a state-of-the-art approach for solving MDPs. Instead of relying on a predefined model, RL maps environmental states to optimal actions through interaction with the environment, which may be either the real system or a simulation of it. RL is characterized by three key properties: (1) a closed-loop process, where the agent's actions influence future states and rewards; (2) the absence of explicit instructions, meaning the model learns purely from feedback rather than predefined rules; and (3) the delayed impact of actions, where decisions affect not only immediate rewards but also long-term outcomes (Sutton & Barto, 2018). RL optimizes decision policies based on sampled experiences $(s, a, r, s')$ as it is learning, rather than relying on full knowledge of the environment. Unlike traditional DP, which calculates Q-values by considering all possible future states and rewards (expected update), RL updates/learns Q-values incrementally with each interaction with the environment (sample-based update). This makes RL particularly advantageous when a complete model of the environment is unavailable or computationally intractable.

The RL process is fundamentally driven by three components: Observation of current state, selection of actions, and pursuit of a defined goal. These repeated interactions form a trial-and-error learning process, where the agent refines its decision-making strategy over time. The interaction loop between the RL agent and the environment is shown in Figure 2. At each discrete time step $t$, the agent observes the current state of the environment $s \in S$, where $S$ is the set of possible states, and selects an action $a \in A(s)$, where $A(s)$ is the set of actions available in state $s$. As a consequence of this action, the environment transitions to a new state $s'$ and provides the agent with an immediate numerical reward $r \in R \subseteq \mathbb{R}$. This iterative process continues, enabling the agent to learn an optimal policy through experience and reward-based feedback.

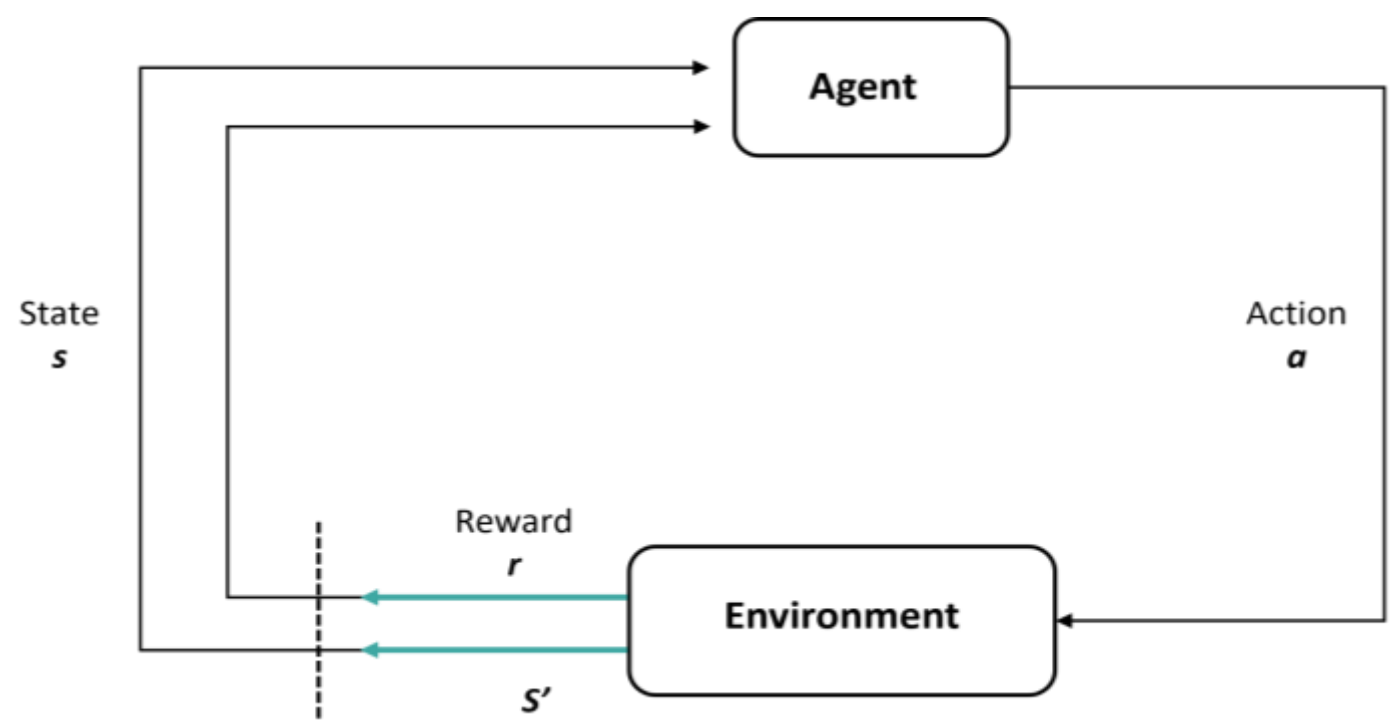

**Figure 2.** Agent-environment interaction loop in Reinforcement Learning.

# Deep Q Network (DQN)

The most widely used method and a fundamental concept in RL is Temporal Difference (TD) learning (Boyan, 2002), which combines elements of both Monte Carlo (MC) methods and Dynamic Programming (DP). Like MC, TD learning updates estimates based on actual experiences without requiring full knowledge of the environment. Similar to DP, it estimates and updates the value of the current state based on estimated values of subsequent states, without waiting until the end of an episode.

Q-learning is an off-policy TD learning algorithm that separates the behavior policy (used for action selection) from the target policy (used for learning). It updates the action-value function using the Bellman optimality equation and uses an ε-greedy policy for balancing exploration and exploitation by choosing a random action with probability ε and the best-known action with probability 1- ε. The Q-learning update rule is given by Equation 22.

$$Q\ (s, a\ ) \leftarrow Q\ (s, a\ )\ + \alpha\ \ [r\ + \gamma\ max_{a'} Q\ (s', a'\ ) - Q\ (s, a\ )] \qquad \text{Eq. } 22$$

Immediately after transitioning to the next state, $s^{'}$, the agent receives a better estimate $r + \gamma\ max_{a'} Q\ (s', a'\ )$ than its old one $Q\ (s, a\ )$. Q-learning uses this difference, $[r\ + \gamma\ max_{a'}\ Q\ (s', a'\ ) - Q\ (s, a\ )]$ , also known as TD error, the first part of which $(r\ + \gamma\ max_{a'}\ Q\ (s', a'\ ))$ acts as the "target" toward which the Q-value is adjusted as the agent interacts with the environment. $\alpha$ is the learning rate, which ranges between 0 and 1, determining the degree of the influence of the TD error on old estimates.

In high-dimensional state-action spaces, the applicability of traditional RL becomes limited as it tries to learn action-value function, for each state-action pair it visits, which leads to high computational costs, large storage requirements, and incomplete state-action coverage during the interactions. To address this, Mnih et al. (2015) developed DQN method, a novel RL agent, which combines deep neural networks with Q-learning, and can efficiently learn optimal policies in high dimensional state-action spaces.

The deep neural networks in DQN approximate the Q-function, replacing the need to store a separate Q-value for every state-action pair. The neural network takes the state as input and predicts Q-values for all possible actions, by parameterizing an approximate function. Using a nonlinear function approximator such as neural networks to approximate the Q values, causes instability and divergence issues in learning (Baird, 1995). The instability is mainly caused by the high correlation between the sequence of observations, used as the training data, since each state-action pair depends on previous interactions. Second, small updates to Q-values can significantly change the policy, causing the agent to explore different parts of the state space unpredictably, leading to unstable learning. Third, because Q-values are updated using their own estimates, errors can compound over time, making convergence difficult.

To address the instability, DQN introduced two key improvements. First, experience replay was implemented to break the correlation between consecutive observations. Instead of updating Q-values immediately after each interaction, experiences $(s, a, r, s^{'})$ are stored in a replay buffer, $U$, and training samples are randomly drawn in batches of the stored data, $b \subset U$, ensuring more independent and diverse updates. Second, a target network is introduced to stabilize learning. Rather than using the same network to predict both Q-values and targets, a separate target network is maintained and updated periodically from the main network. This reduces the correlation between the predicted and target Q-values, leading to more stable learning.

After each interaction step, $i$, the Q network uses a loss function, $L_i\ (\Theta_i)$, defined according to the TD error, to update the Q values:

$$L_i(\theta_i) = \mathbb{E}_b\left[\left(r\ + \gamma\ max_{a'}\ Q\ (s', a'; \theta_i^-\ ) - Q\ (s, a; \theta_i)\right)^2\right] \qquad \text{Eq. } 23$$

$\theta_i$ are the parameters (weights) of the Q network at iteration $i$, and $\theta_i^-$ are the parameters of the target network used to compute the target values. In the original DQN, these are copied from the main Q-network at regular intervals and remain fixed in between; in our implementation they are updated gradually using a soft update, as described in the next section.

# Formulating the Problem in a RL Framework

This section describes our formulation of the sequential investment decision problem in a RL framework.

## State Space

The states in our framework consist of several variables that represent the status of the environment, providing useful information for the agent to make informed investment decisions. These variables include:

- Number of the year in a 25-year horizon, $i$
- Oil and gas production of a year (million BOE/year), $Prod_{oil,i}$
- Electricity production from renewable resources of a year (GWh), $Prod_{rnw,i}$
- Installed renewable capacities of a year (GW), $Capacity_{rnw,i}$
- $CO_2$ emissions of a year (million tons), $Emissions_{CO_2,i}$
- Social Costs of Carbon of a year (USD/tons $CO_2$ emission), $SCC_i$
- Oil price of a year (USD/barrel), $Price_{oil,i}$
- Electricity price of a year (million USD/GWh), $Price_{electricity,i}$
- Previous investment shares $Inv_{oil,i}$ , $Inv_{rnw,i}$, $Inv_{CO_2,i}$ (fractions summing up to 1)
- Available budget for investment (million USD), $B_i$
- And renewable Competitive Advantage index, $CA_i$

All state variables are min–max normalized to [0,1] using fixed bounds before being passed to the network.

## Action Space

The action space consists of investment allocations across the three sectors, oil and gas, renewables, and CCS. Investment shares are discretized in increments of 0.05 from 0 to 1, resulting in 231 distinct allocation combinations, each summing to 1.

We use a discrete action formulation deliberately: strategic allocations are set in coarse increments in practice, and the resulting 231-action set makes value-based DQN directly applicable, with interpretable Q-values per allocation alternative.

## Reward Structure

Clear objectives and trade-offs – a fundamental element in good decision making (Howard & Abbas, 2015)– are defined by a well-structured reward function in RL. The reward function must be carefully designed so that maximizing the reward corresponds to achieving the decision objectives – maximizing profit, reducing the social costs of carbon, and enhancing the company's competitive advantage in renewables.

Multiple objectives in RL adds complexity to the learning process, as policies must balance competing rewards rather than optimizing a single objective (Roijers et al., 2013). In Q-learning, this challenge is further amplified since the agent must optimize a vector-valued Q-function instead of a scalar (Van Moffaert & Nowé, 2014). This leads to difficulties in action selection, as no single optimal policy exists, but rather a set of Pareto-optimal solutions (Vamplew et al., 2011). Moreover, conflicting objectives can cause unstable Q-value updates, resulting in slower convergence (Hayes et al., 2022). The exploration-exploitation trade-off also becomes more complex, as improving one objective may impact another (Mannion et al., 2017).

To ease the process, we combine the three objectives into a single reward function with multiple criteria. As explained in the Simulation Environment section, all three objectives are measured in monetary terms, making them easy to compare and combine. To balance their importance, we use the swing weighting method (Keeney & Raiffa, 1993), which assigns weights based on how much each objective from its worst to best value impacts the overall value of the decision. The final weights are $w_{Profit}$ = 0.55, $w_{SCC}$=0.2 and $w_{CA}$=0.25, for maximizing profit, reducing SCC, and enhancing CA, respectively. Each value is min–max normalized to [0,1] before weighting, using fixed bounds corresponding to the objectives' extreme attainable values. For the social cost of carbon, the normalization is oriented so that lower costs map to higher values, and all three terms are maximized.

The combined reward function includes two additional components – penalties – to account for the constraints: one for net-zero emission regulations, $r_{NetZero}$, and the other for smooth transition constraints, $r_{smoothness}$, as explained in previous sections. The value for these penalties is 0.25 per violated milestone and per sector whose allocation change reaches 0.15 and are determined through trial and error in such a way that they do not dominate other reward components or become dominated by them. The combined reward function for calculating reward in each interaction step, $r_i$, is:

$$r_i = w_{Profit} * NPV_i + w_{SCC} * SCC_i + w_{CA} * CA_i + r_{NetZero} + r_{smoothness} \quad \text{Eq. 24}$$

## DQN Architecture and Integrated Diagram

The schematic representation of the neural network in our DQN model is shown in Figure 3. This network takes the current state as input and outputs Q-values for all possible actions. We implement this architecture using the Python library PyTorch (Paszke et al., 2019). Our model consists of three fully connected hidden layers with Rectified Linear Unit (ReLU) activation functions. The first hidden layer contains 256 units, the second has 128 units, and the third has 64 units. The final output layer has the same number of nodes as the action space, 231, corresponding to the Q-values for each possible action in the given state.

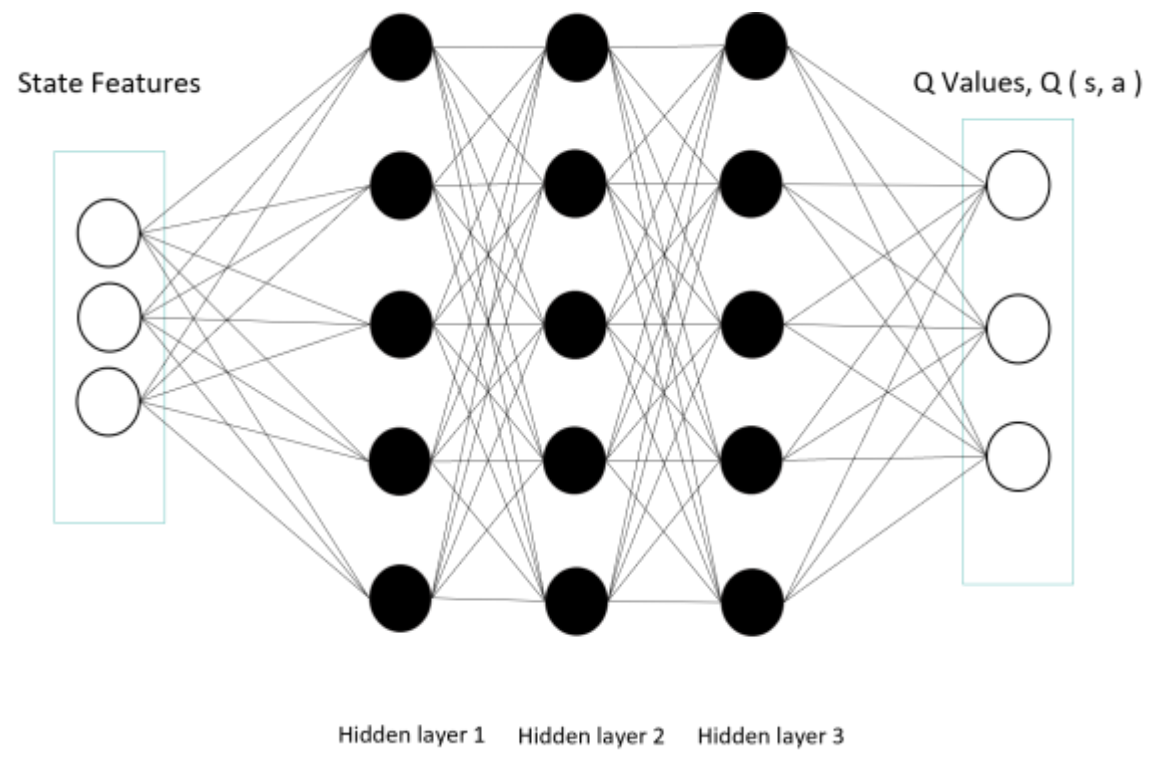

**Figure 3.** Schematic representation of the neural network in our DQN model.

Figure 4 shows the training process of the DQN algorithm and the interaction between the RL agent and the environment. The process begins by initializing the environment with the starting conditions, including the initial investment budget, energy production levels, and market variables as described in the "Simulation Environment" section. The Q-network weights are randomly initialized, and an experience replay buffer is created to store past interactions.

At each time step, the agent observes the current state, $s$, and selects an action based on the ε-greedy policy. The agent selects a random action with probability ε (exploration), or the action with the highest Q-value from the main Q-network with probability 1-ε (exploitation). The value of ε starts at 0.999 ensuring pure exploration initially, and decays at a rate of 0.99985 per episode to gradually shift toward exploitation. A minimum ε value of 0.01 is maintained to allow a possibility of exploration after convergence. The selected action, $a$, determines the investment allocation across the three sectors for that year.

Once the action is taken, the environment updates according to the predefined transition models and returns the next state, $s'$, and the immediate reward calculated from the single reward function (Eq. 24). Here, an episode corresponds to one complete simulated project trajectory over the full planning horizon, while each step represents a single decision time step (year). The experience $(s, a, r, s')$ is then stored in the experience replay buffer with a capacity of 100,000 experiences, and older experiences are removed as new ones are created. Instead of updating the Q-network immediately after each experience, a random batch of 256 past experiences are sampled from the replay buffer. The TD error is then computed using the target network and is used to adjust the weights of the main Q-network using the loss function, as shown in Eq. 23. The target network is updated using a soft update scheme at each step, where its parameters are gradually adjusted toward those of the main Q-network. This interaction loop continues for 30,000 episodes until the total reward stabilizes, indicating convergence to an optimal policy.

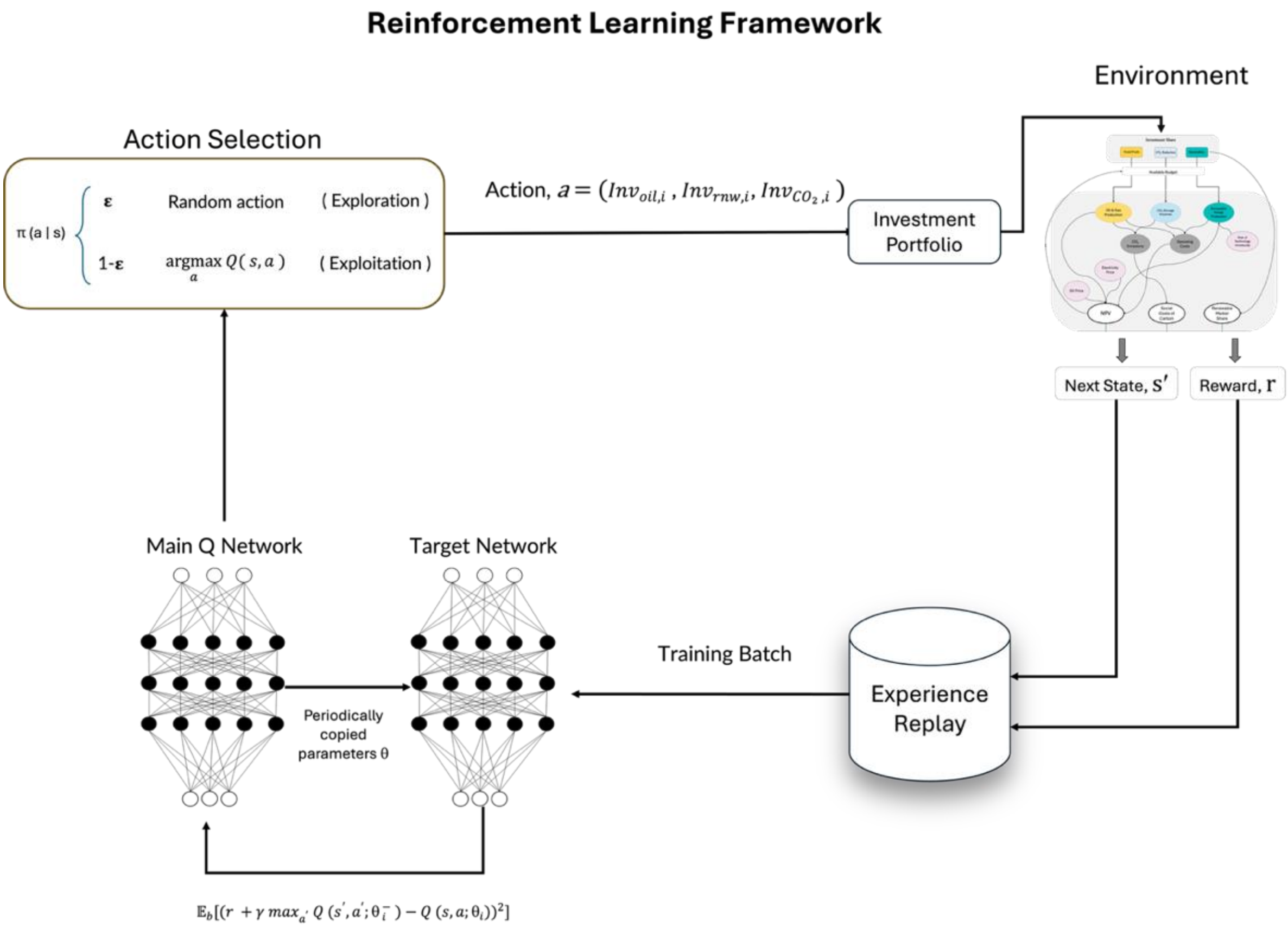

**Figure 4.** Training cycle of the DQN algorithm and the interaction between the RL agent and the environment.

## Implementation Details and Training

The model is implemented in Python using PyTorch, with the environment structured using the OpenAI Gym interface (Brockman et al., 2016). Training is performed on a system with an Intel Core i9 CPU and an NVIDIA RTX A2000 GPU to accelerate computation and reduce training time. The model was trained over 30,000 episodes using the Adam optimizer (Kingma & Ba, 2014), with a learning rate of $\alpha = 0.00005$ and a discount factor of $\gamma = 0.95$. The training performance is shown in Figure 5, where the vertical axis shows the total reward obtained in each episode, and the horizontal axis indicates the episode number. The training is well performed as the total reward converges to around 10.8. The reward curve fluctuates slightly even after convergence due to the uncertainties in the environment and the residual exploration. The value reported in Table 1 corresponds to the best policy identified during training, evaluated in pure exploitation mode over the 1,000 realizations.

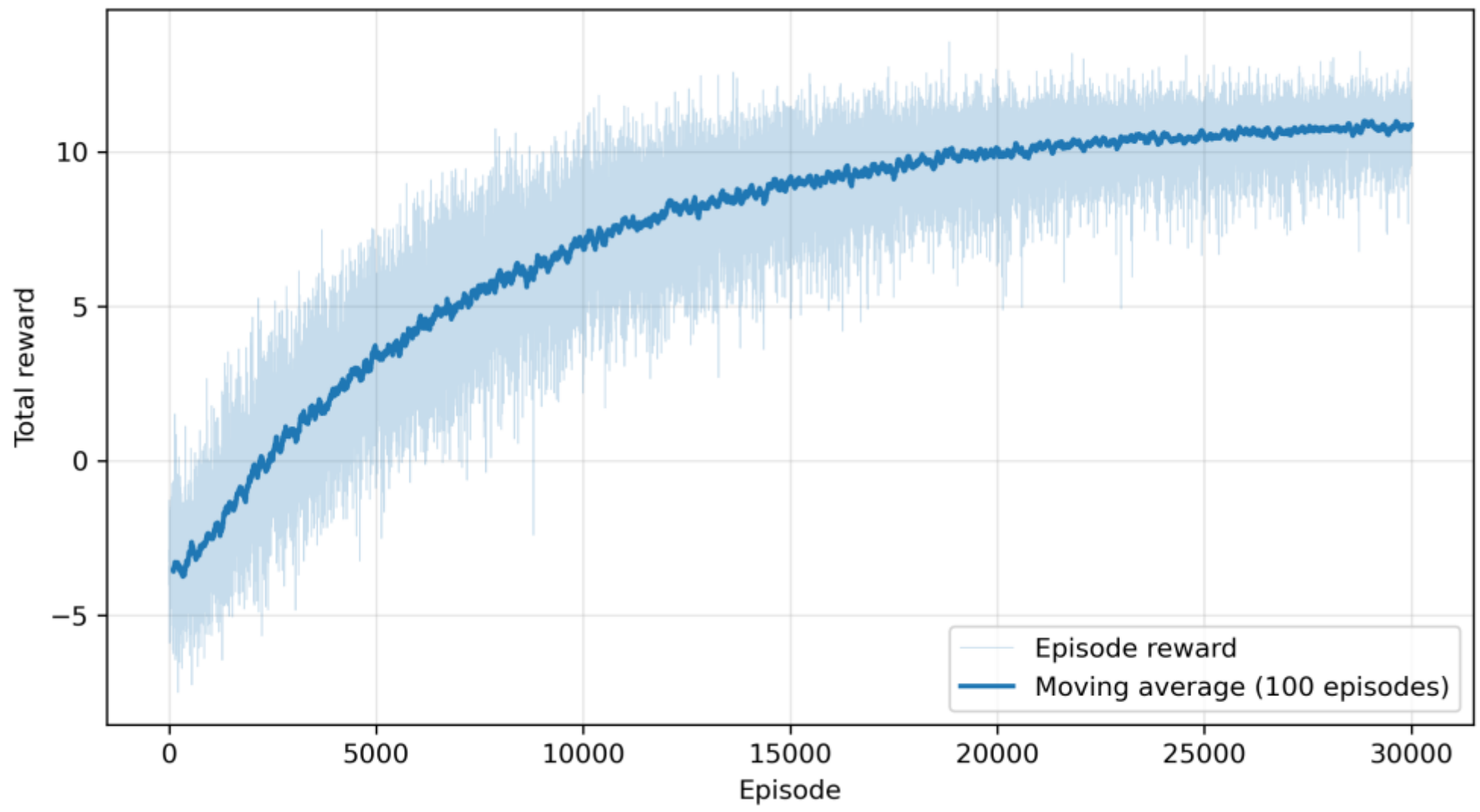

**Figure 5.** Training performance of the DQN algorithm over episodes.

# Results and Discussion

In this section, we present the results of the sequential investment decisions for the simulated environment, using the RL framework. To evaluate the optimality of the RL's obtained policy, we compare the results with several benchmark strategies as will be described. We also present Pareto plots to illustrate the trade-offs between objectives, showing how the improvement in one objective impacts others.

## RL Optimal Policy

Figure 6 shows the average investment allocations over 25 years, solved by RL. The average investment allocations are based on 1,000 episodes of interaction with the environment, following the obtained optimal policy in pure exploitation mode (no exploration). These 1000 episodes have been generated using fixed random seeds and are used for all subsequent analysis and evaluations.

In the early years, starting from a base investment budget of 1 billion USD, most of the funds are allocated to the oil and gas sector, with around 70% of the total investment. The RL agent learns this behavior to take advantage of higher short-term returns in oil and gas and to generate capital for future investments in renewables and CCS.

In later years, investments in oil and gas become less attractive due to rising operating costs and the progressively tightening net-zero emission constraints, as modeled in the environment. Renewables are also funded from the beginning, with allocations generally around 20–30%. Early investment in renewables is essential in this framework to maximize competitive advantage, by positioning the decision-maker as an early adopter. This objective encourages early participation in the renewable market and is captured by the RL agent, despite no immediate reward being assigned to renewable investments in the first five years.

Over time, investment allocations in oil and gas decline, while renewable investments share gradually increases and becomes dominant in the later years, exceeding 60% of the annual budget. CCS investments begin in the early–mid period and stabilize around 30%. The allocation to CCS is strongly influenced by $CO_2$ emissions and the regulatory constraints imposed to achieve net-zero targets. The slight increase in CCS investment toward the end of the horizon reflects the need to meet the final emission reduction targets and avoid penalties. The detailed behaviors may not be fully realistic and are influenced by the simplified assumptions of the simulation environment, which may deviate from real-world dynamics.

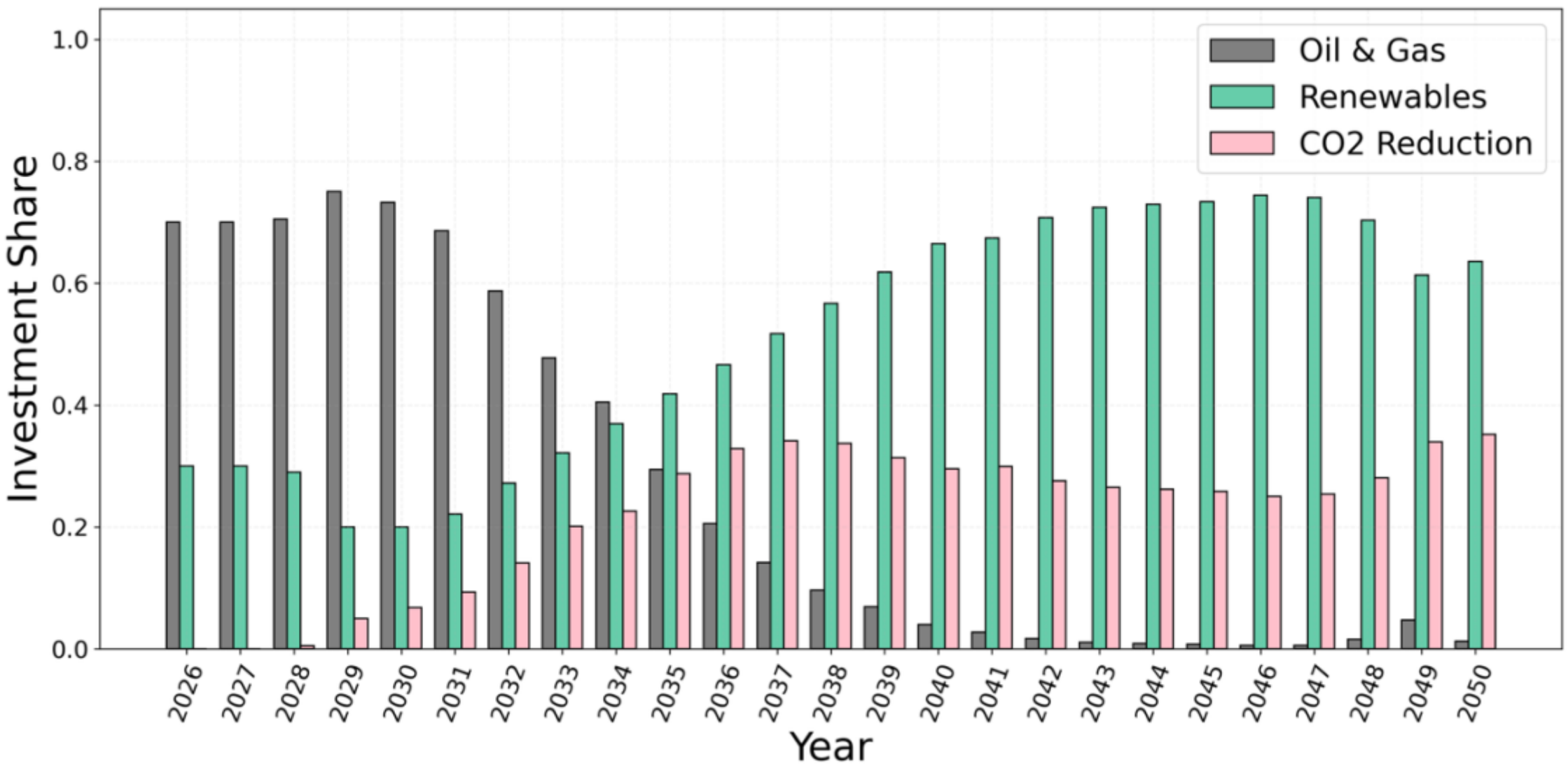

**Figure 6.** Averages of optimal investment allocations suggested by RL.

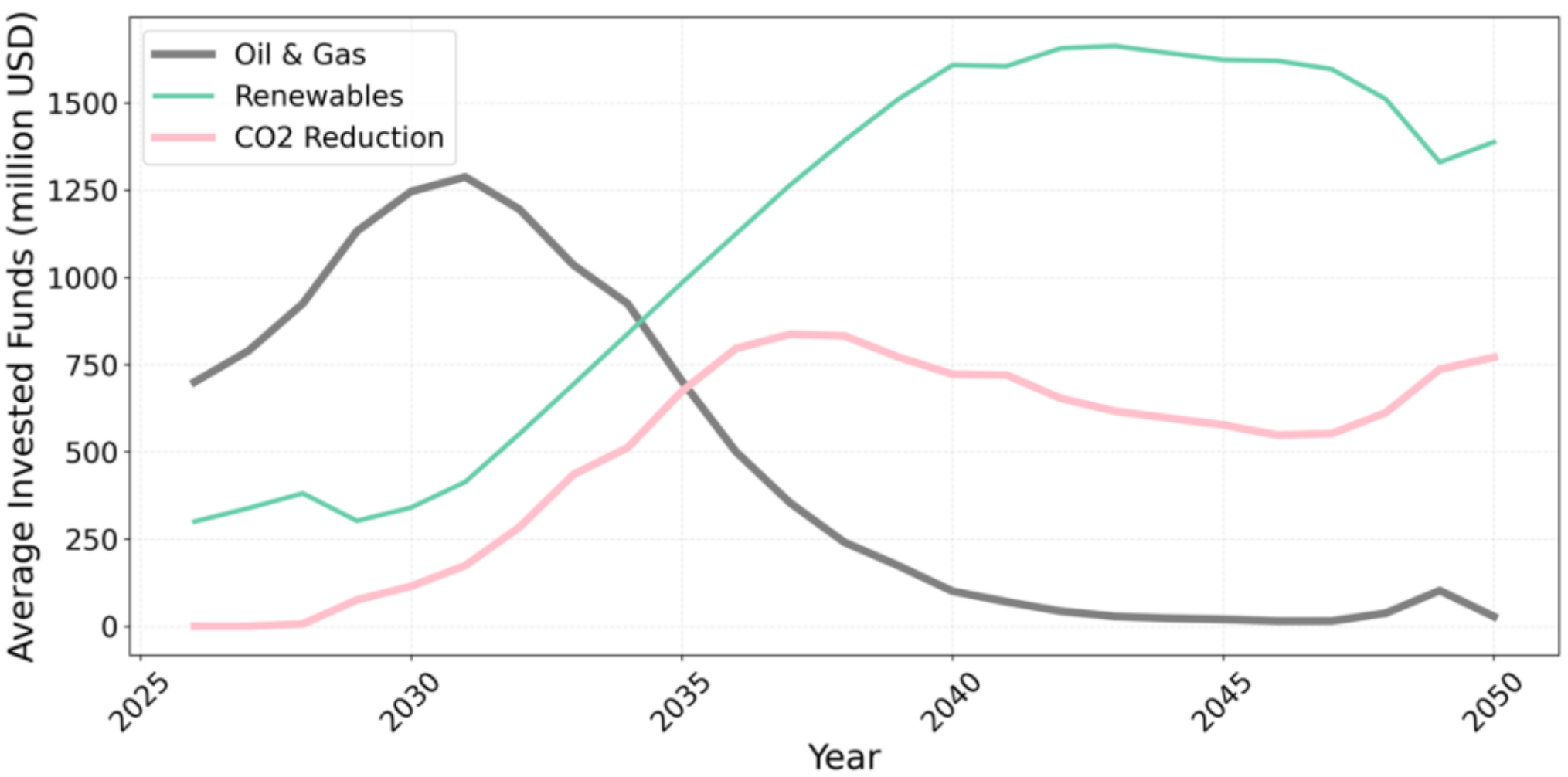

**Figure 7.** Actual annual investment in each sector.

Multiplying the allocation shares by the available annual budget, Figure 7 shows the actual funds invested in each sector over time. As will be shown, the available budget increases toward the middle of the planning horizon and then gradually declines toward the end, reflecting the shift toward less profitable sectors. It is also worth noting that, in this framework, we assume the entire available budget is to be allocated across the three sectors. In practice, however, it may be optimal not to invest the full budget in certain periods.

Due to the uncertainties, the selected investment shares can vary across episodes, even under a fixed decision policy. Figure 8 shows the P10–P90 ranges of investment shares, respectively for oil and gas, renewables, and $CO_2$ reduction, over the 1,000 episodes used for analyzing the optimal policy suggested by RL.
These ranges reflect how the RL agent responds to different states arising from uncertainty in the environment. For a given state, the policy itself is deterministic and produces a consistent decision; however, variability in the state trajectories leads to variability in the resulting investment allocations across episodes. Although uncertainty could be better characterized with more detailed and realistic models, some level of variability in decisions is inherent in stochastic environments.

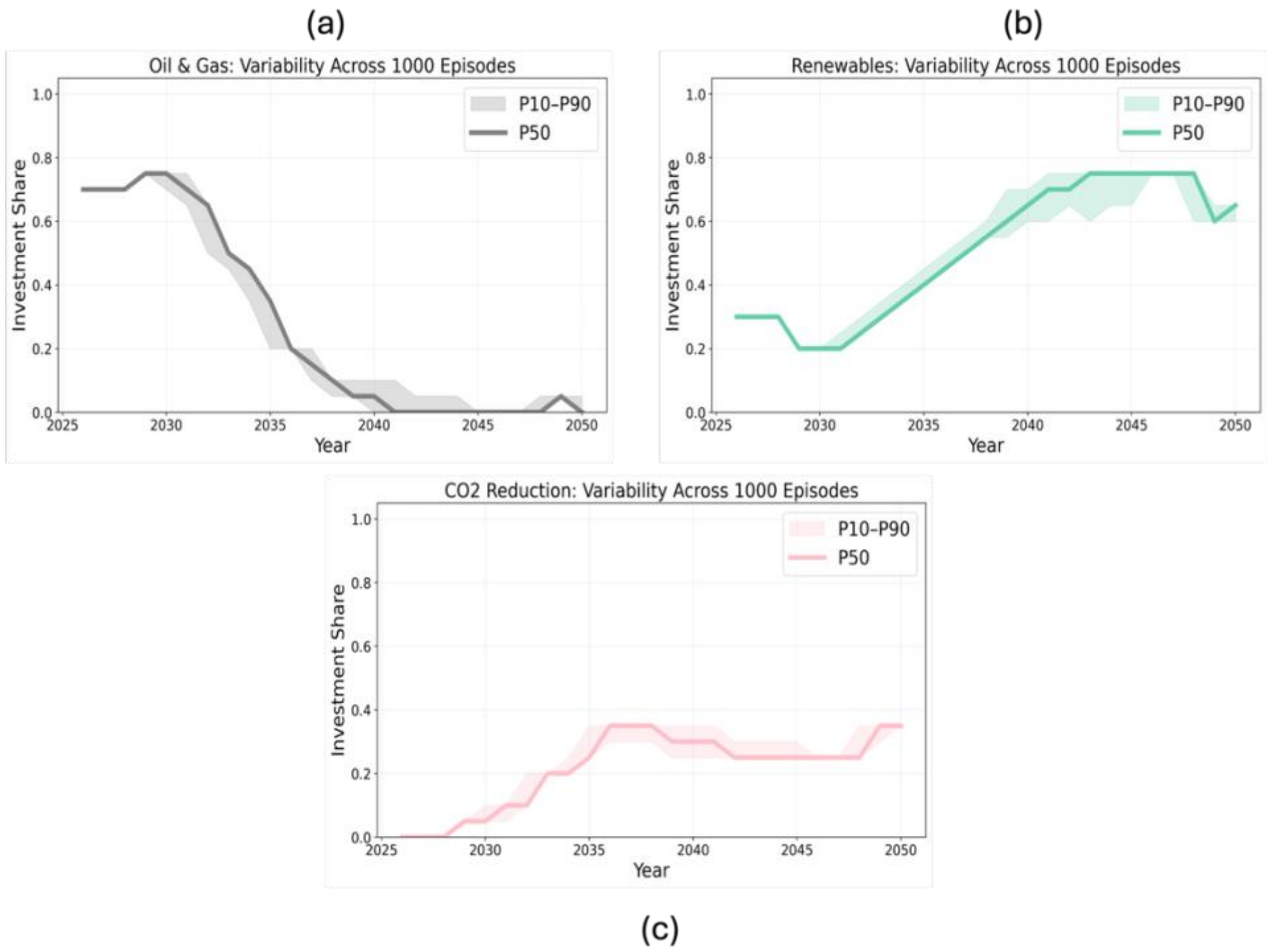

**Figure 8.** P10-P90 ranges of investment shares for (a) oil and gas, (b) renewables, and (c) $CO_2$ reduction sectors, over the 1,000 episodes based on the optimal policy from RL.

In addition to investment decisions, our framework also provides insight into the impact of the decisions on oil and gas production, renewable energy production, $CO_2$ emissions and economic performance throughout the transition period, as shown in Figure 9.

Oil production slightly increases in the early years and then starts to fall after reaching a peak around 109 million barrels of oil equivalent around year 2033. This pattern follows the early focus on oil and gas investments, which later decrease as the policy responds to tightening emissions constraints and less profitability of this sector. After 2033, production drops quickly and goes below 26 million barrels by 2050.

Renewable energy production shows strong and steady growth, especially after 2030. Average annual output rises from about 800 GWh in 2026, which is well below the installed capacity's potential, reflecting the high probability of underperformance events in the early years, to about 27,000 GWh by 2050. This rise is directly linked to the larger share of investment going into renewable projects with technology improvement and as the decision policy responds to tightening $CO_2$ emissions constraints and the associated benefits of renewables in enhancing competitive advantage. The variability in renewable production increases during the transition period but reduces noticeably after 2045. This is consistent with the technology immaturity risk formulation in the model, where the risk of underperformance diminishes over time and is assumed to be zero after this year. As a result, investments translate more directly into effective capacity, leading to a sharper and more consistent increase in renewable output across scenarios.

The $CO_2$ emission profile closely follows the trend of oil and gas production, which is the only source of emissions in this model. Emissions begin at about 29 million tons and remain relatively stable in the early years, peaking at 30 million tons around 2031. After 2033, emissions decline sharply as investments shift away from oil and gas toward CCS and renewables. By 2040, average emissions fall to about 13 million tons, and they continue to decrease toward zero by 2050. The learned policy successfully satisfies the imposed emission reduction targets, achieving reductions of approximately 24%, 58%, 81%, and 100% by 2035, 2040, 2045, and 2050 respectively, exceeding all defined milestone thresholds. Similarly, $CO_2$ intensity, defined as emissions per unit of energy production, follows a declining trend throughout the whole transition period.

Cash flow starts high, averaging about 5.2 billion USD per year from 2026 through 2033, because oil and gas projects give strong returns early on. It then falls to a minimum of roughly 1.8 billion USD per year in 2043 as the focus moves to renewables and CCS, which cost more upfront and take longer to pay off. After that, cash flow recovers to about 3.5 billion USD by 2050, driven by the accumulation of installed renewable capacity and the reduction in technology immaturity risk in the later years, as also reflected in the renewable production profile.

The available budget increases in the early years and peaks at about 2.5 billion USD around 2038, supported by strong operating cash flows from oil and gas activities. As the budget is partly determined by reinvestment of operating cash flow, this early profitability enables a rapid expansion in investment capacity. After the peak, the budget declines moderately, ending at about 2.2 billion USD in 2050, reflecting the lower short-term returns and higher upfront costs of the sectors that dominate the later period. The smoother decline in the later years is influenced by the budgeting structure, which incorporates a memory effect that dampens short-term fluctuations in cash flow.

## Comparison with Benchmark Decision Policies

One aim of this study is to evaluate whether RL can identify a near-optimal investment decision policy in the face of long-term uncertainty and competing objectives. We compare the performance of the RL-derived policy against a set of intuitively simple decision rules. All benchmark strategies are evaluated under the identical 1000 realizations. The benchmark decision rules are:

1- **Greedy Policy**: Investment decisions are selected to maximize immediate reward, without consideration for long-term impacts.
2- **Equal Allocation**: Investments are distributed approximately equally across oil & gas, renewables, and $CO_2$ reduction in every year, subject to discretization of the action space.
3- **Early Transition to Renewables**: A swift transition strategy in which investment is shifted toward renewables from the beginning. Specifically, from 2025 to 2035, 70% of the annual budget is allocated to renewables and 30% to oil & gas. From 2035 to 2045, the renewable share increases to 85% and in the final five years, 100% of the budget is allocated to renewables. No investment is allocated to $CO_2$ reduction in this strategy.
4- **Late Transition to Renewables**: A delayed transition strategy in which investment in renewables is postponed until later in the planning horizon. From 2025 to 2040, the entire budget is allocated to oil & gas. Starting in 2040, 50% of the budget is allocated to renewables and 50% to oil & gas. This share increases to 100% allocation to renewables in the final years of the horizon. No investment is allocated to $CO_2$ reduction in this strategy.
5- **Single-Focus Strategies**: Three additional baselines where all available funds are allocated exclusively to a single sector throughout the planning horizon.

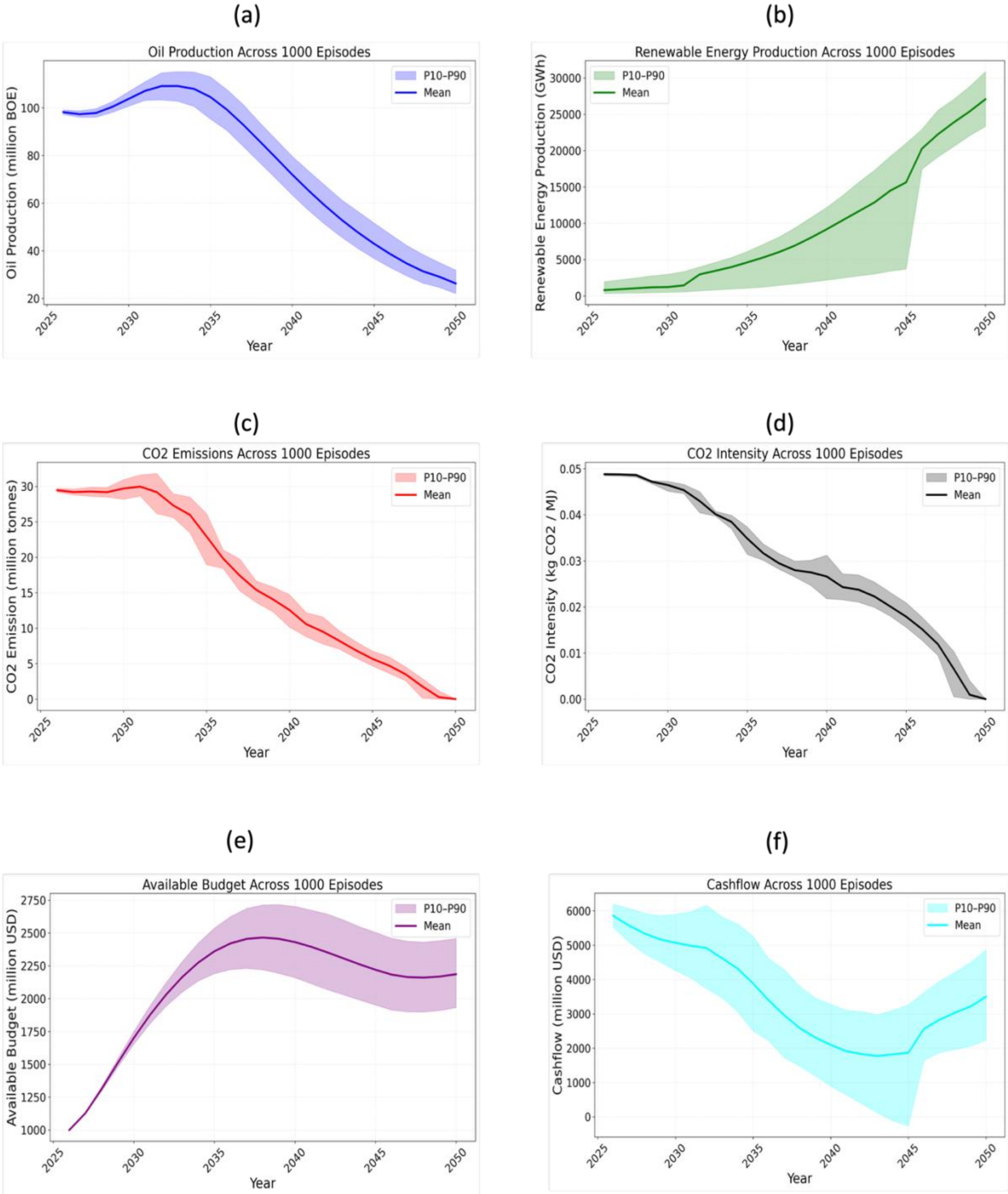

**Figure 9.** Projected trends of key variables during the energy transition under the optimal policy, showing mean values and P10–P90 uncertainty ranges across 1,000 episodes. (a) Annual oil production, (b) Annual renewable energy production, (c) Annual $CO_2$ emissions, (d) Annual $CO_2$ intensity, (e) Available investment budget, (f) Annual cashflow.

We emphasize that these benchmarks are not intended to reflect realistic, real-world policies, but rather serve to illustrate how RL performs relative to a diverse set of alternatives.

As the objective is to maximize expected cumulative reward, we evaluate the performance of each decision policy based on the cumulative reward averaged across 1,000 Monte Carlo simulated uncertainty paths over the 25-year horizon. The results are summarized in Table 1. In addition to the combined reward (Eq. 24), we also show the three individual objective rewards. The applied constraints are incorporated into the combined reward formulation. Based on the combined reward which integrates all three objectives, RL's strategy clearly outperforms the others. When focusing only on maximizing profit, the strategy that allocates entirely to the oil and gas sector achieves the highest reward. Similarly, for the $CO_2$ reduction, the strategy that invests only in $CO_2$ reduction methods performs best. It is worth noting that the renewables-only strategy does not achieve the highest competitive advantage reward. This is due to the formulation of the CA metric, which rewards not only high investment levels but also the timing and gradual increase of investment in renewables over time. The greedy, equal, early and late transition strategies lie between these extremes, with the equal allocation strategy gaining the highest combined reward among them.

**Table 1** Comparison of RL against several defined benchmarks.

| Strategy | Combined Reward | Profit Reward | $CO_2$ Reduction Reward | CA Reward |
|---|---|---|---|---|
| RL | 11.26 | 9.85 | 14.75 | 11.63 |
| Greedy | 8.78 | 9.46 | 14.37 | 3.93 |
| Equal | 9.38 | 8.59 | 15.08 | 8.55 |
| Early Transition | 8.93 | 9.59 | 14.05 | 8.94 |
| Late Transition | 6.82 | 15.49 | 1.53 | 0.0 |
| Only Oil & Gas | 8.14 | 16.35 | 0.74 | 0.0 |
| Only Renewable Energies | 9.91 | 7.9 | 18.38 | 8.55 |
| Only $CO_2$ Reduction | 6.69 | 3.41 | 24.05 | 0.0 |

## Multi-Objective Trade-offs and Pareto Front Analysis

Understanding the impact of trade-offs between conflicting objectives on decisions and outcomes is essential in decision-making. Pareto analysis provides useful insight by exploring these trade-offs and showing how improving performance on one objective can affect the others.

For the Pareto front analysis, we trained the DQN agent fifteen times, each with a different set of weights assigned to the three objectives in the reward function (Eq. 24). The weights vary in steps of 0.25 and sum to one. The best policy obtained for each weight configuration was saved and subsequently evaluated over the 1000 realizations. The average values of the objectives were then used to construct three pairwise plots in Figure 10, illustrating the trade-off space.

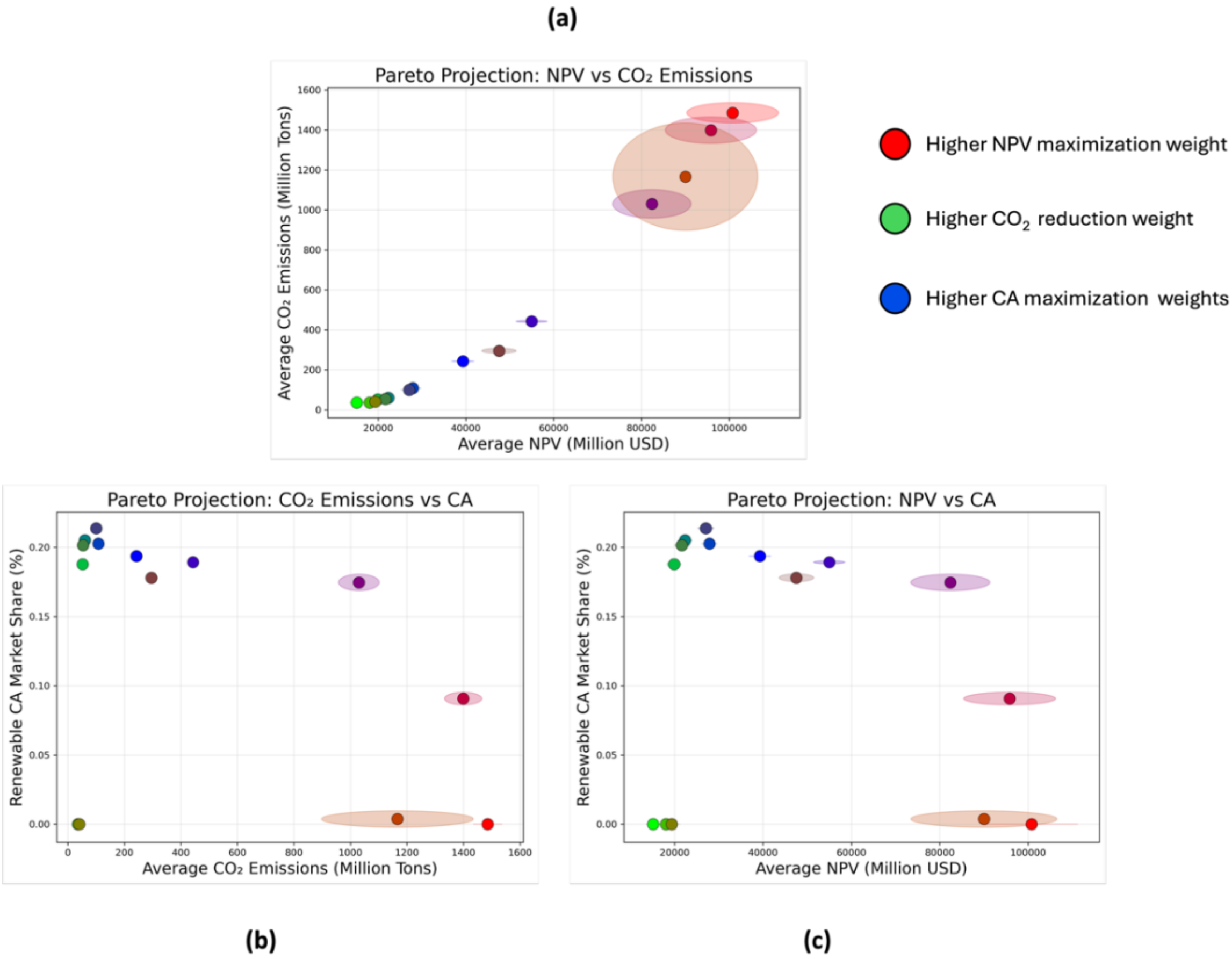

**Figure 10.** Pareto front visualizations showing the impact of trade-offs between (a) NPV and $CO_2$ emissions, (b) CA and $CO_2$ emissions, and (c) NPV and CA under different objective weight settings. Colors indicate the dominant training objective.

Figure 10a shows the trade-off between NPV (representing the objective of profit maximization) and total $CO_2$ emissions (representing the objective of $CO_2$ social cost reduction). Figure 10b presents the relationship between $CO_2$ emissions and CA, while Figure 10c compares NPV with CA. Each color represents the dominant objective during training: red indicates high weight in profit maximization, green for high $CO_2$ cost reduction weight, and blue for high CA weight. The ellipses reflect the uncertainty of outcomes for each weight configuration. Each data point shows the cumulative NPV and $CO_2$ emissions over the 25-year period, together with the final-year CA market share, averaged across the simulated realizations.

Figure 10a shows a clear Pareto front: higher NPV accompanied by higher emissions in general. The front is not fully linear. On average, reducing one million tons of $CO_2$ sacrifices roughly 57 million USD in NPV, but this cost is not the same everywhere. When emissions are high, reducing them by one million tons costs about 44 million USD in NPV. When emissions are already low, the same reduction costs about 110 million USD. The cheapest reductions therefore come first, and the last steps toward net-zero are the most expensive. Trade-offs in Figures 10b and 10c are more complex due to the influence of the third objective. The presence of multiple objectives can make the relationship between any two objectives dependent on others.

Figure 10b shows two main patterns in the relationship between $CO_2$ emissions and CA. Focusing first on the low-emission region, when the weight on profit is low or zero, introducing weight on CA (from greener to more blue points) raises the competitive advantage substantially, from zero to about 0.19%, with a modest increase in emissions. Beyond that point, further weight on CA yields little additional advantage, as CA plateaus around 0.20%, while emissions continue to rise. It is interesting that policies with a dominant weight on $CO_2$ reduction appear at both low and high CA levels. This suggests that, in the current environment, emissions reduction and CA can be aligned, depending on how investments are allocated. In contrast, when the weights shift between CA and NPV while the $CO_2$ weight remains low, reducing the emphasis on CA leads to a sharp drop in CA and a clear increase in emissions. The observed relationship between $CO_2$ emissions and CA depends strongly on the third objective.

A similar pattern is seen in Figure 10c for CA vs. NPV. When the weights are mainly on emissions reduction and CA, both objectives increase together, because investment in renewables contributes to both $CO_2$ reduction and profitability. However, after CA reaches its maximum, further weight placed on profit leads to a significant increase in NPV at the cost of reduced CA.

These results emphasize that the interactions among objectives are complex in a multi-objective setting and trade-off analysis must account for how all objectives interact. Focusing on only two metrics may give an incomplete picture of the interactions.

# Decision Quality of the Proposed Framework

In this section, we assess the quality of the proposed decision framework using the Decision Quality (DQ) chain (Howard & Abbas, 2015), which consists of six elements that together characterize the quality of a decision process. Figure 11 shows this chain and maps each element to its specific counterpart in our study, and the following discussion evaluates each element in turn. To summarize the assessment, we assign an indicative score to each link, reflecting how well it is fulfilled within the scope of this study.

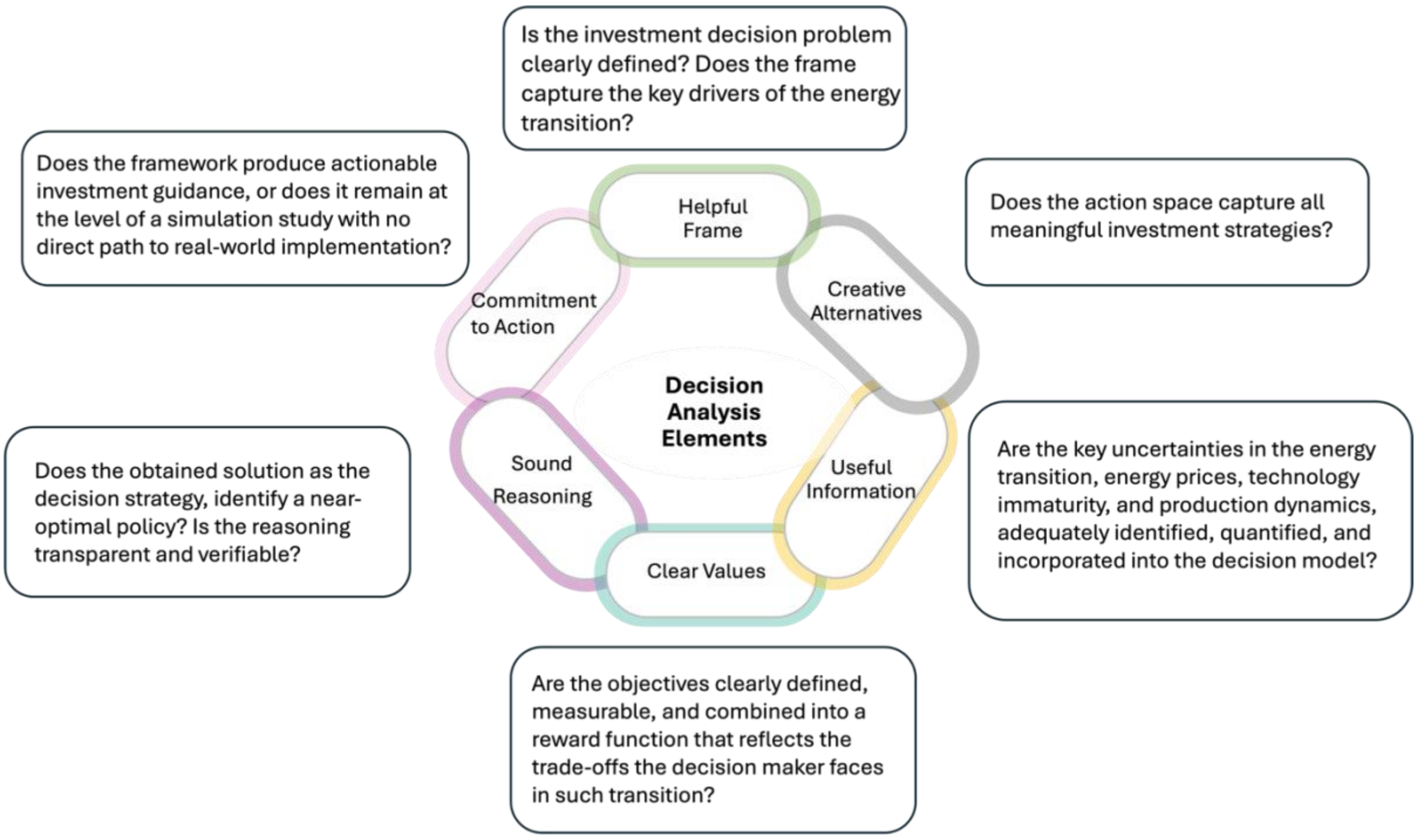

**Figure 11.** Decision quality chain of the proposed framework.

1. **Helpful Frame (80%)** requires that the decision problem be clearly defined to the decision maker. In this study, the frame is explicit: an energy company allocates investments annually across three sectors over a 25-year horizon, balancing three competing objectives under uncertainty. The scope of the simulation environment, including what is modeled and what is excluded, is transparently communicated. However, one limitation is that the frame was defined entirely by the authors rather than in collaboration with a real decision maker, meaning that some of the framing choices, such as the sectors included and the length of the horizon, were driven by what could be practically modeled rather than by the priorities of an actual company.
2. **Creative Alternatives (50%)** is arguably the weakest link in the DQ chain for this study. The action space spans only 231 discrete allocation combinations, and several benchmark policies are evaluated for comparison. However, the alternatives considered are constrained, since the decision is always framed as how to distribute the full available budget across the same three sectors each year. For example, different strategic alternatives, such as withholding investment in unfavorable periods, divesting assets, or allocating funds to other sectors, such as research and development, are not in the current action space and considering more creative alternatives would change the obtained policy and the allocation of investments.
3. **Useful Information (60%)** requires that key uncertainties are identified, quantified, and incorporated into the decision model. In this study, the main sources of uncertainty which are oil and electricity prices, technology immaturity risk, and production variability, are explicitly modeled using stochastic processes. The model assumptions are transparently described and formulated; however, the uncertainty representation is simplified. For example, price dynamics follow GBM assumption with fixed parameters, production variability is captured through simple Gaussian distributions rather than data-driven models and the technology immaturity risk is based on intuitive assumptions with parameters that are not supported by a verified study or model. In addition, other important uncertainties such as regulatory changes, energy demand, and competitive dynamics are not included. The information used is therefore sufficient to support the study's illustrative purpose, but a real decision maker would require richer and more context-specific information to act on these insights.

4. **Clear Values (85%)** requires that the decision maker's objectives and trade-offs are explicitly defined. This is a strong link in this study. Three objectives are defined and all converted to monetary units to allow direct comparison. Trade-offs between objectives are made explicit through assigned weights, and the sensitivity of outcomes to different weight combinations is explored through Pareto analysis. These values are clearly structured and honestly communicated, but their relevance depends on how well they represent the priorities of a real energy company.
5. **Sound Reasoning (80%)** requires that the solution method is appropriate and the results are transparent and verifiable. In this study, the problem is formulated as a Markov Decision Process and solved using a Deep Q-Network, a well-established method for SDM under uncertainty. The training process is documented, convergence is verified, and the policy is benchmarked against several alternatives. The optimality of the RL policy holds within the boundaries of the simulation environment; if these boundaries are changed, the agent must be retrained to obtain the corresponding optimal policy.
6. **Commitment to Action (40%)** refers to the transition from decision making to actual implementation. As a simulation-based research framework, it produces an optimal investment policy but stops short of real-world implementation. No actual decision maker is involved, and no concrete steps toward execution are defined. This is appropriate for a research study whose primary goal is to demonstrate the value of the proposed framework and methodology. However, bridging the gap between the derived policy and actionable guidance for a real energy company would require collaboration with actual decision makers, more specific modeling assumptions, and a clear implementation roadmap.

# Conclusions and Future Work

This study introduced a novel decision-support framework to evaluate and optimize long-term investment strategies for energy companies navigating the energy transition. We framed the optimization of investment strategy as a Sequential Decision-Making problem under uncertainty and developed an illustrative and representative simulation environment modeling the dynamic energy landscape up to 2050. Reinforcement Learning, specifically a Deep Q-Network, was applied to learn optimal investment strategies that optimize multiple and conflicting objectives (profitability, emissions reduction, and competitive advantage in renewables) for given trade-offs. The optimality of the RL-derived strategy was tested against several benchmark strategies. In addition, the Pareto front analysis was conducted to understand the impact of trade-offs among competing objectives on investment outcomes.

Our main conclusion is that framing the energy transition as a multi-objective sequential decision making under uncertainty is essential for better investment decisions and value creation; the value here is not only profit but also the value from $CO_2$ reduction and competitive advantage as quantified by the combined reward in our example. Traditional sequential decision optimizers like Dynamic Programming approaches are limited in handling high-dimensional problems. In contrast, Reinforcement Learning learns optimal policies directly from simulated experiences, without requiring prior knowledge of the environment. Through repeated interactions between the RL agent and modeled environment, the RL agent can efficiently and effectively solve for a (near-)optimal decision policy balancing immediate and long-term value creation.

As expected, the RL policy outperformed benchmark strategies, achieving the highest combined reward with a smooth change in allocation of investments over time. Under the RL policy, the shift from oil and gas to renewables is gradual but accelerates after 2033, aligning with $CO_2$ reduction targets while preserving financial stability. Net-zero emissions are achieved by 2050, but this comes at a high cost: annual cash flow declines through the 2030s to a minimum in the early 2040s before recovering as renewable capacity matures, which highlights the financial trade-offs that companies must plan for. Such insight emphasizes the importance of advanced, long-term planning in navigating the energy transition with both strategic and financial foresight.

Several areas of our simulation environment and modeling approach could be further improved. For example, the environment can be made more realistic by incorporating additional critical variables such as energy demand, which would allow for the representation of supply-demand dynamics, and the scope can be expanded to include other sectors, such as research and development to enrich the decision-making portfolio. Moreover, other material uncertain variables can be included. This study employs simple Gaussian distributions to represent uncertainty in variables like energy production. Future research can leverage historical data to build more context-specific stochastic models. Last but not the least, while our current reward structure combines multiple objectives into a single scalar function for tractability, future work can explore other multi-objective reinforcement learning approaches aimed at identifying Pareto-optimal policies. These approaches may better capture the inherent trade-offs between competing objectives, though it remains a relatively underdeveloped area in current literature and toolsets.

In conclusion, our work demonstrates that combining Sequential Decision-Making thinking with modern Reinforcement Learning techniques offers a powerful toolset for tackling the complex, uncertain challenges in energy transition investments. The proposed framework, which is openly available, provides a flexible and accessible foundation for future research on decision-making in this high-stakes domain.